\documentclass[fleqn,usenatbib,useAMS]{mnras}
\usepackage{graphicx}
\usepackage{amsmath}
\usepackage{amssymb}
\usepackage{multicol}
\usepackage{pdflscape}
\usepackage[T1]{fontenc}
\usepackage{ae,aecompl}
\usepackage{newtxtext,newtxmath}
\usepackage{listings}
\usepackage{xcolor}
\usepackage{ulem}

\title{Mapping gravity in stellar nurseries -- establishing the effectiveness of 2D acceleration maps}

\author[He et al.]{
Zhen-Zhen He$^{1, 2}$, 
Guang-Xing Li$^{3}$\thanks{Contact e-mail: \href{mailto:gxli@ynu.edu.cn}{gxli@ynu.edu.cn}, \href{mailto:ligx.ngc7293@gmail.com}{ligx.ngc7293@gmail.com}},
and Andreas Burkert$^{4,5}$\\
$^{1}$Department of Astronomy, Yunnan University, Kunming, China, \\
$^{2}$Research Center for Intelligent Computing Platforms, Zhejiang Laboratory, Hangzhou 311100, China\\
$^{3}$South-Western Institute For Astronomy Research, Yunnan University, Kunming, China\\
$^{4}$ University Observatory Munich, Scheinerstrasse 1, D-81679 M\"unchen, Germany\\
$^{5}$ Max-Planck-Fellow, Max-Planck-Institute for Extraterrestrial Physics, Giessenbachstrasse 1, 85758 Garching, Germany
}

\begin{document}
\maketitle

\begin{abstract}
    Gravity is the driving force of star formation.
    Although gravity is caused by the presence of matter, its role in complex regions is still unsettled. 
    One effective way to study the pattern of gravity is to compute the accretion it exerts on the gas by providing gravitational acceleration maps.
    A practical way to study acceleration is by computing it using 2D surface density maps, yet whether these maps are accurate remains uncertain.
    Using numerical simulations, we confirm that the accuracy of the acceleration maps $\mathbf a_{\rm 2D}(x,y)$ computed from 2D surface density are good representations for the mean acceleration weighted by mass.
    Due to the under-estimations of the distances from projected maps, the magnitudes of accelerations will be over-estimated $|\mathbf a_{\rm 2D}(x,y)| \approx 2.3 \pm 1.8 \; |\mathbf a_{\rm 3D}^{\rm proj}(x,y)|$, where $\mathbf a_{\rm 3D}^{\rm proj}(x,y)$ is mass-weighted projected gravitational acceleration, yet $\mathbf a_{\rm 2D}(x,y)$ and $ \mathbf a_{\rm 3D}^{\rm proj}(x,y)$ stay aligned within 20$^{\circ}$.
    Significant deviations only occur in regions where multiple structures are present along the line of sight.
    The acceleration maps estimated from surface density provide good descriptions of the projection of 3D acceleration fields.
    We expect this technique useful in establishing the link between cloud morphology and star formation, and in understanding the link between gravity and other processes such as the magnetic field.
    A version of the code for calculating surface density gravitational potential is available at \url{https://github.com/zhenzhen-research/phi_2d}.
\end{abstract}
l


\section{Introduction} \label{sec:intro}
Stars form in the dense parts of the molecular interstellar medium (ISM) \citep{2007ARA&A..45..481Z,2014prpl.conf....3D}, whose evolution is controlled by a variety of physical processes: turbulence \citep{2004ApJ...615L..45H,2011ApJ...740..120R}, gravity \citep{2011MNRAS.411...65B,2012MNRAS.427.2562B}, and magnetic fields \citep{2015MNRAS.452.2410S}.
Gravity is a fundamental, long-range force that plays a decisive role in the evolution of molecular clouds.

The gravitation acceleration map provides a new view into the star formation processes \citep{2016arXiv160305720L}.
Gravity originates from the mere presence of matter which can lead to further concentrations, yet the spatial distribution of mass affects how gravity acts.
One interesting case is that matter tends to gather at the edge of sheets and at the tips of filaments \citep{2007ApJ...654..988H, 2015MNRAS.449.1819C}.
One crucial step in understanding gravity is to produce maps of the gravitational field, measured in terms of acceleration for the observed clouds.

An observational limitation is that gas can be reliably traced on the sky plane, where only 2D surface density can be derived.
One connivent way to study acceleration is to compute it using the surface density.
This approach was called by \citet{2016arXiv160305720L} as ``gravitational acceleration mapping''.
However, the accuracy of such 2D acceleration maps remains a question.
Using numerical simulations, we study the relations of acceleration computed from the 2D surface to the real 3D volume density.
In the past, the gravitational potential map may be used to find cores.
During the formation of cores from diffuse matter, gravitational potential deepen during accretion.
\citet{2011ApJ...729..120G} calculate the gravitational potential maps using volume density ($\Phi$) and surface density ($\Phi_{\rm2D}$) respectively.
They identify regions within the largest closed gravitational potential isosurface as a core.
Based on their result, there is not much difference in core-finding between the two maps. We believe that the value of the projected gravitational potential is to study the effect of gravitational acceleration on cloud evolution, hence this investigation.

\section{Data}\label{sec:data}
The simulation data is taken from the StarFormMapper project\footnote{https://starformmapper.org/home}. We use the results from the ``Barotropic EOS cluster simulations" performed using the AREPO moving-mesh code \citep{2010MNRAS.401..791S,2019MNRAS.486.4622C}. The density field is presented in an SPH-like format, which is projected onto a rectangular grid using the cloud-in-cell (CIC) algorithm, which is essentially a smoothing with kernels of adaptive sizes.

\begin{figure*}
\centering
\includegraphics[width=.4\textwidth]{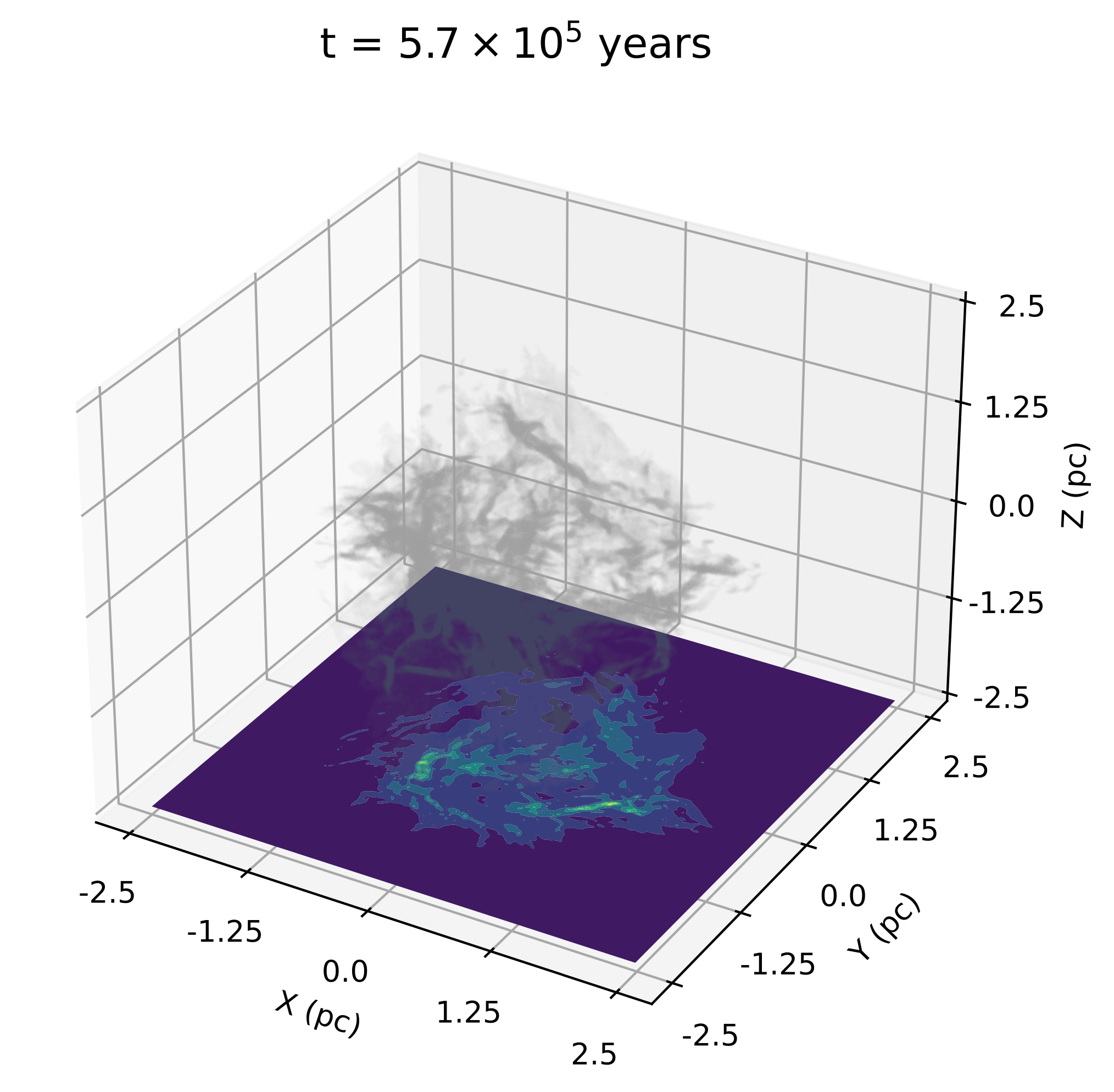}
\includegraphics[width=.4\textwidth]{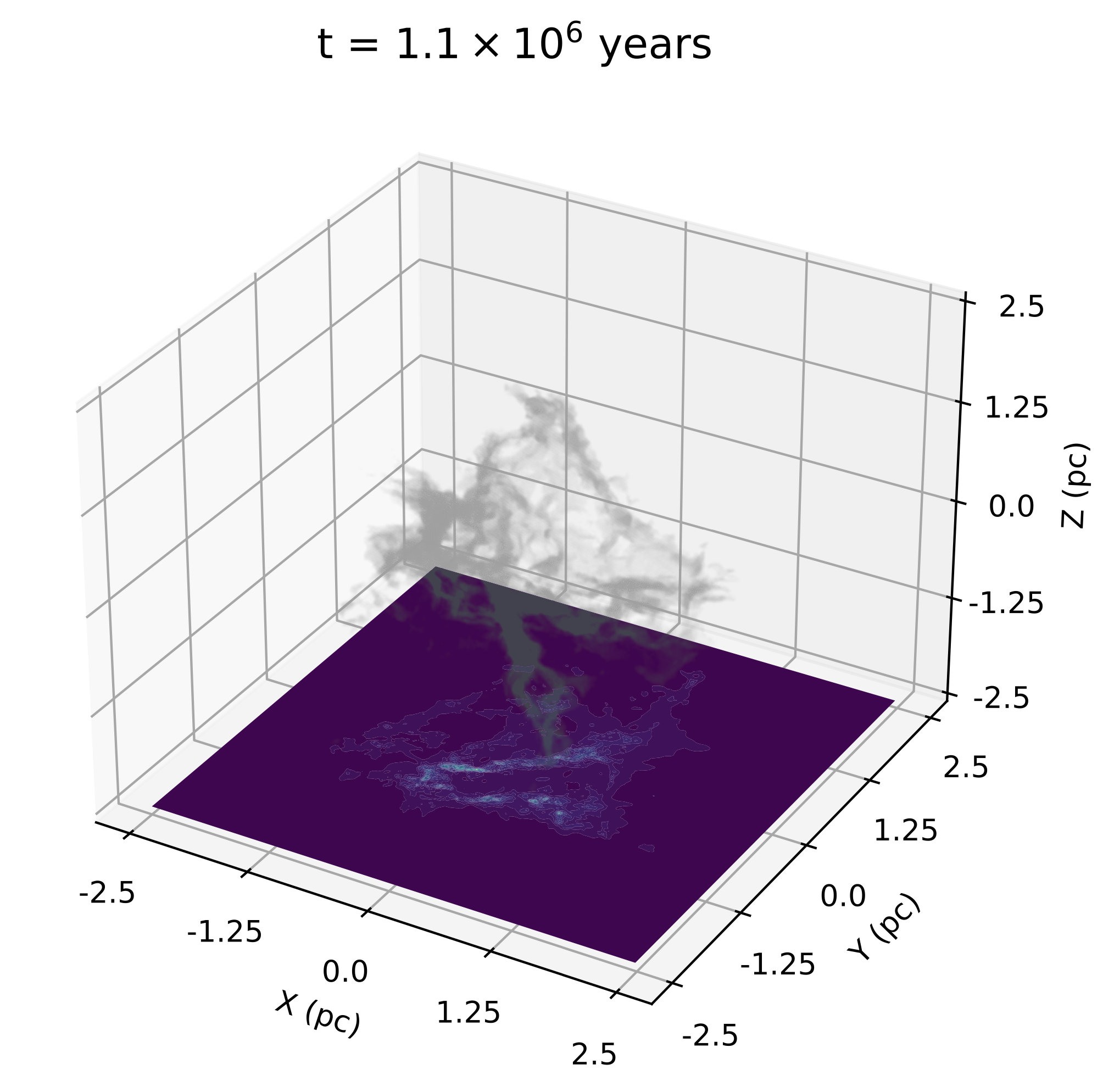}
\caption{The simulated 3D data is taken from the StarFormMapper project. Each cloud has a total mass of about 1000 M$_\odot$.
Left panel shows cloud at time of $\sim5.7 \times 10 ^ 5$ years and right panel shows cloud at time of $\sim1.1 \times 10 ^ 6$ years. The 2D surface density map was generated by integrating the 3D density along the $z$ axis.}\label{fig:data_scatter}
\end{figure*}

We take two cloud snapshots at different evolutionary stages.
The simulated clouds (see Fig. \ref{fig:data_scatter}) have a total mass of about 1000 M$_\odot$, and most of the gas concentrates in a cubic space with sides of $\sim 5\;\rm pc$.
The physical parameter resembles a typical star-forming region such as the NGC1333 \citep{2006AJ....131.2921R}.
The one at $t = 5.7 \times 10^5$ years captures the moment where the turbulence has created a network of dense filamentary structures whose collapse just began, and in the one $t = 1.1 \times 10^6$ years, the collapse has already progressed significantly.
Our aim is to study the accuracy of the acceleration mapping method for structures at different evolutionary stages.
 
To simulate the observed 2D surface density, we project 3D volume onto a 2D plane to generate a surface density map $\Sigma(x,y)$ by integrating the density along the $z$ axis.

\section{Calculating gravitational accelerations} \label{sec:cal}
To derive the acceleration, we first compute the gravitational potential based on the density distribution.
The gravitational potential is computed through Poisson's equation
\begin{equation}
\nabla^2 \Phi = 4\pi G\rho\;,
\end{equation}
where $\Phi$ is the gravitational potential, $G$ is the gravitational constant, and $\rho$ is the density of the matter.

In the 3D case, Poisson's equation can be solved efficiently in the Fourier space:
\begin{equation}
\Phi_{k, \rm 3D} = -\frac{4\pi G\rho_k}{k_{\rm 3D}^2}\;.
\end{equation}
In the case of 2D, assuming 3D density is distributed in a thin plate of half-thickness $H$, the potential is \citep{2011ApJ...729..120G}
\begin{equation}\label{eq:2d}
\Phi_{k, \rm 2D} = -\frac{2\pi G\Sigma_k}{|k_{\rm 2D}|(1+|k_{\rm 2D}H|)}\;.
\end{equation}
$\Phi_k$ is the gravitational potential in the $k$ space, $\rho_k$ and $\Sigma_k$ is volume density and surface density in the $k$ space for the 3D and 2D case, respectively, $k_{\rm 3D} = \sqrt{k_x^2 + k_y^2 + k_z^2}$ for the 3D case, and $k_{\rm 2D} = \sqrt{k_x^2 + k_y^2}$ in the 2D application.

To derive the gravitational potential, one can transform the density
distribution into the $k$ space, compute $\Phi_k$, and back to the real space to gett $\Phi$.
Finally, gravitational acceleration can be derived using
\begin{equation}
\mathbf {a} = \nabla \Phi.
\end{equation}
All the calculations are performed on Cartesian grids.
A Python code for calculating the gravitational potential of a 2D density plane is available at GitHub website \footnote{\url{https://github.com/zhenzhen-research/phi_2d}}.

Using the Fourier method to calculate gravitational potential automatically assumes periodic boundary conditions. This means that the matter in the box repeats itself and extends to infinity periodically. Gravity is a long-range force, the matter outside the box may affect the gravitational potential value inside the box. To reduce the influence of periodicity, we doubled the size of the boxes and put zeros on the expanded regions. Because gravitational acceleration scales as $r^{-2}$ where $r$ is the radius, under sufficient padding, the influence of periodicity is minimum.

\section{Results}\label{sec:results}
We first calculate the gravitational potential and 3D acceleration $\mathbf a_{\rm 3D}(x,y,z)$ using the 3D volume density, as well as the 2D accelerations $\mathbf a_{\rm 2D}(x,y)$ using the 2D surface density map.

\begin{figure*}
\centering
\includegraphics[width=0.4\textwidth]{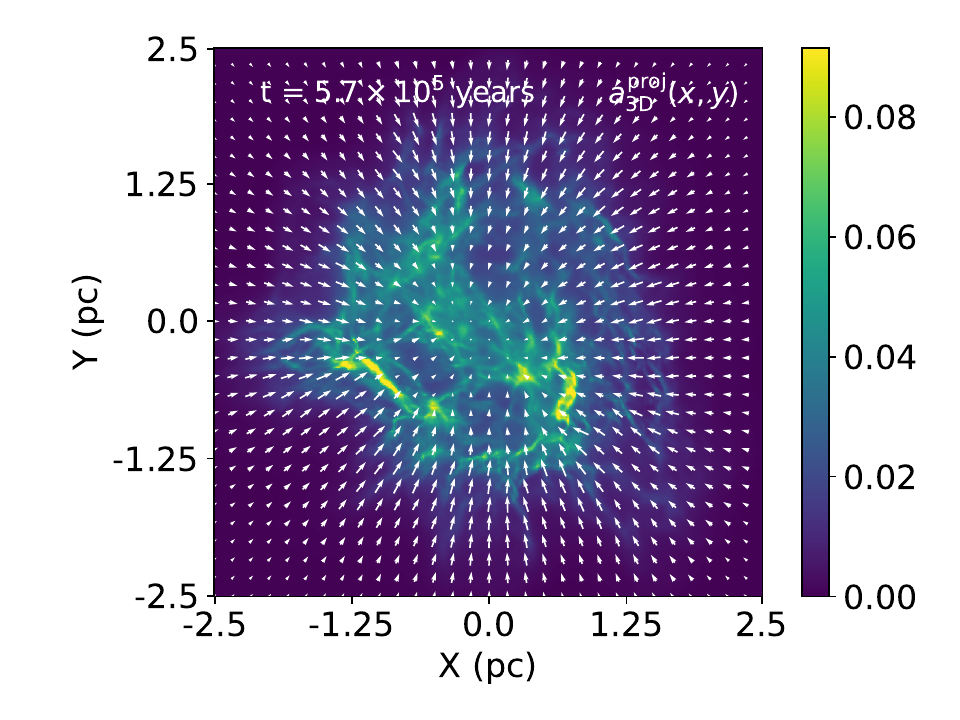}
\includegraphics[width=0.4\textwidth]{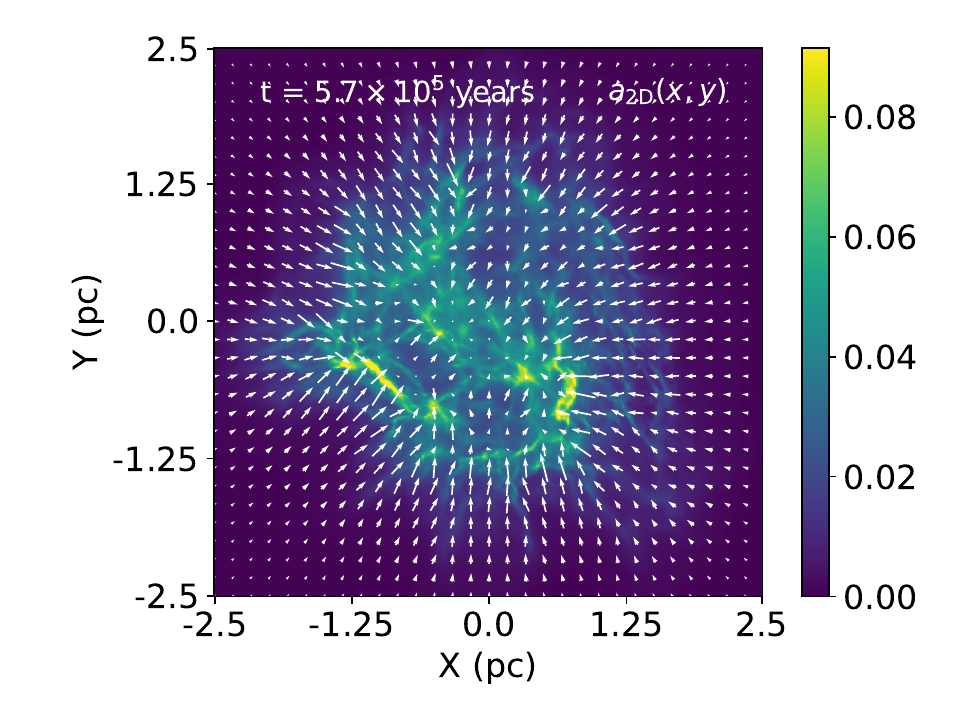}\\
\includegraphics[width=0.4\textwidth]{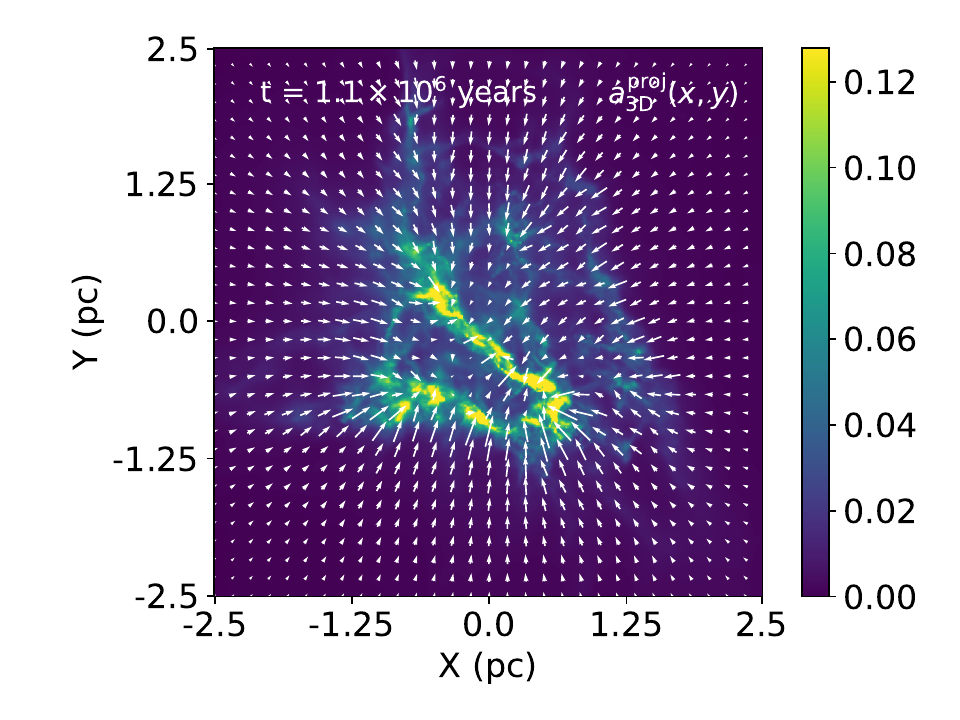}
\includegraphics[width=0.4\textwidth]{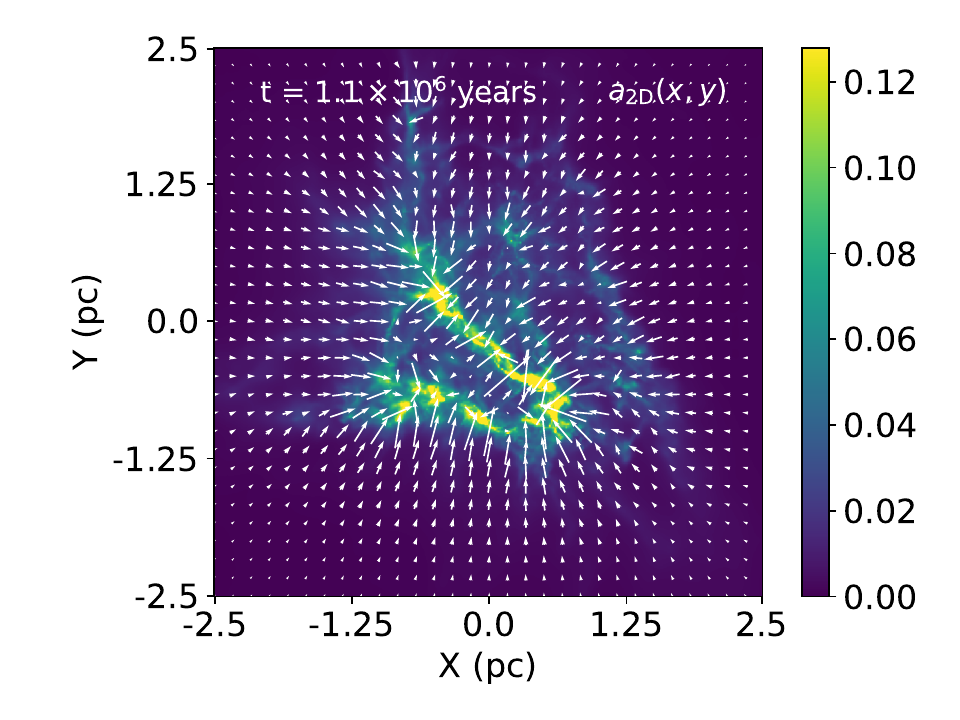}
\caption{Acceleration map $\mathbf a_{\rm 3D}^{\rm proj}(x, y)$ and $\mathbf a_{\rm 2D}(x, y)$ of clouds at different evolutionary stages overplotted onto the density map.
The vectors stand for acceleration and the background image is the density distribution.}\label{fig:2D}
\end{figure*}

We are interested in whether the accelerations along the $x$ and $y$ can be constrained through the 2D calculations.
To facilitate comparisons, we define the projected acceleration $\mathbf a_{\rm 3D}^{\rm proj}(x, y)$ as
\begin{equation} \label{eq}
\mathbf {a}_{\rm 3D}^{\rm proj}(x, y) = \frac{\int \mathbf {a}_{\rm 3D} (x,y,z) \rho(x,y,z){\rm d} z}{\int \rho(x,y,z) {\rm d} z} \;,
\end{equation}
where the acceleration component of $\mathbf {a}_{\rm 3D} (x,y,z)$ along the $z$-axis direction is discarded. The projected acceleration map $\mathbf a_{\rm 3D}^{\rm proj}(x,y)$ as well as the acceleration map $\mathbf a_{\rm 2D}(x,y)$ constructed from projected density are shown in Fig. \ref{fig:2D}.

\subsection{Accuracy of Acceleration map}\label{sec:comp}
\begin{figure*}
\centering
\includegraphics[width=0.35\textwidth]{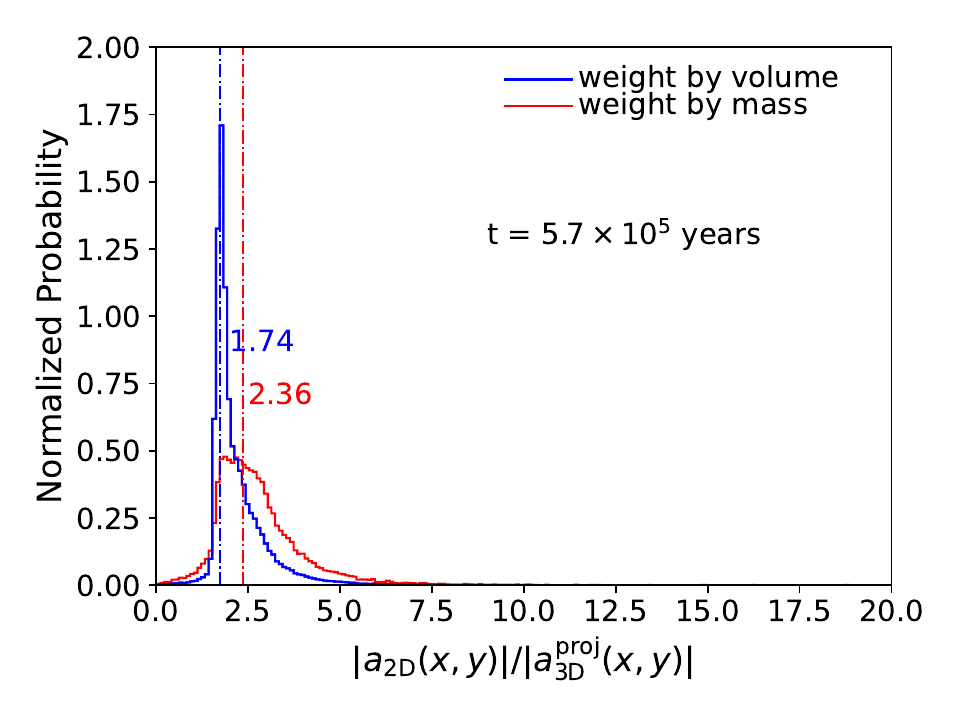}
\includegraphics[width=0.35\textwidth]{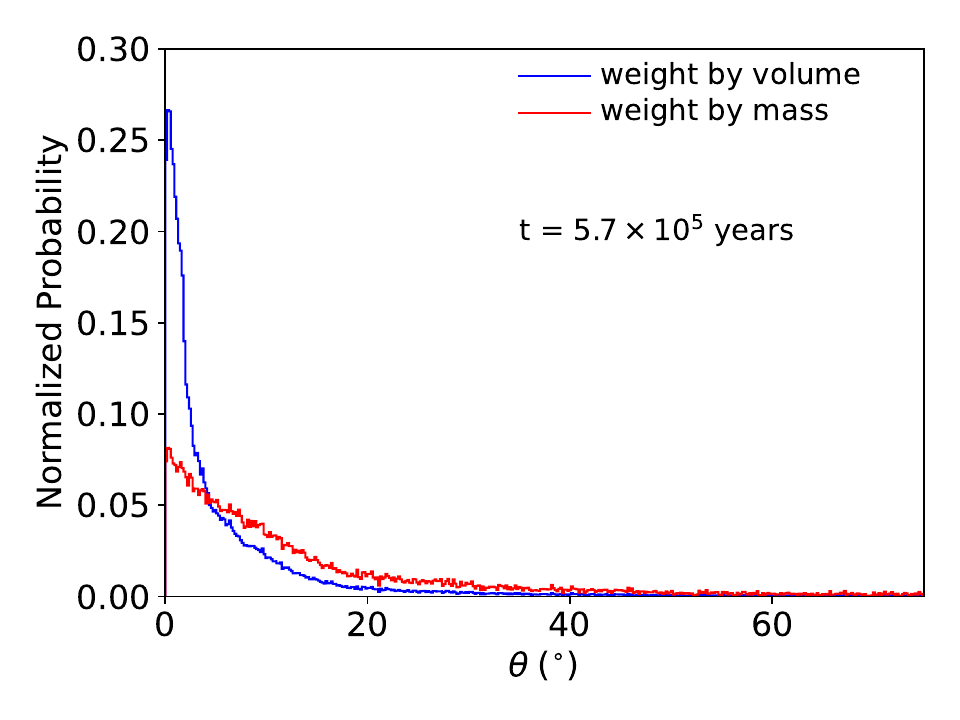}\\
\includegraphics[width=0.35\textwidth]{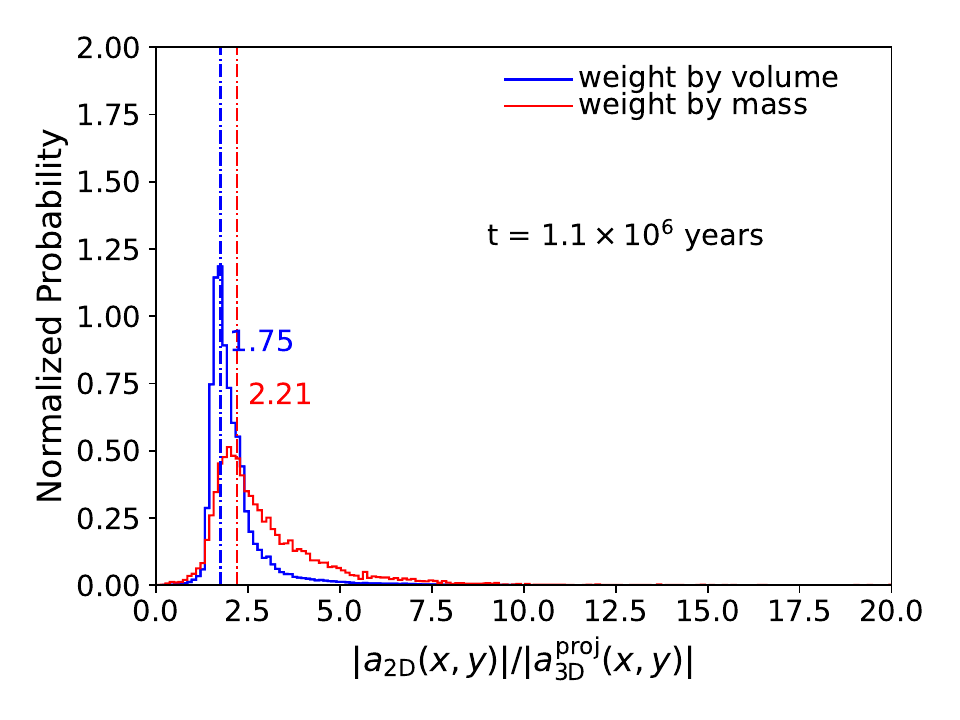}
\includegraphics[width=0.35\textwidth]{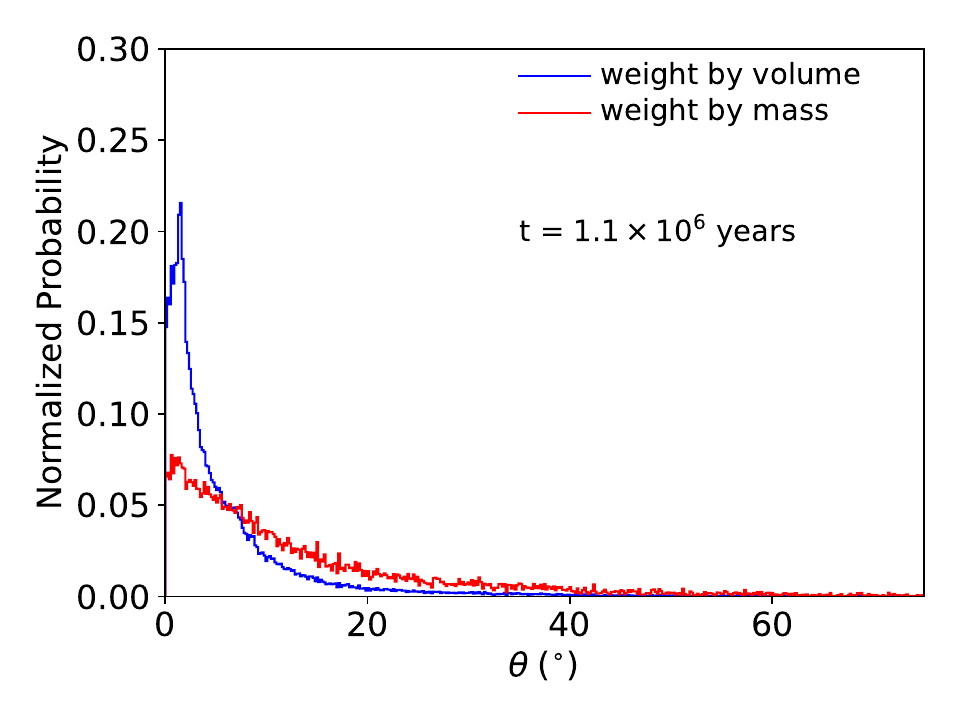}
\caption{The proportion of amplitude ratios ${|\mathbf a_{\rm 2D}(x,y)|/|\mathbf
a_{\rm 3D}^{\rm proj}(x,y)|}$ and include angles $\theta$. The blue lines are the probability
histogram weighted by volume, the red lines are the probability histogram weighted by mass.
The amplitude ratios $|\mathbf a_{\rm 2D}(x,y)| \approx 2.3 \pm 1.8 \; |\mathbf a_{\rm 3D}^{\rm proj}(x,y)|$ which is weighted by mass, the majority of include angles stay within 20$^{\circ}$.}\label{fig:count}
\end{figure*}

To quantify the differences between $\mathbf a_{\rm 3D}^{\rm proj}(x,y)$ and $\mathbf a_{\rm 2D}(x,y)$, we plot the distributions of vector amplitude ratios ${|\mathbf a_{\rm 2D}(x,y)|/|\mathbf a_{\rm 3D}^{\rm proj}(x,y)|}$ as well as included angles $\theta$ between $\mathbf a_{\rm 2D}(x,y)$ and $\mathbf a_{\rm3D}^{\rm proj}(x,y)$ in Fig. \ref{fig:count}, where we plotted both the volume-weighted and the mass-weighed distributions.
The angle between two vectors can be obtained from the inner product: $\mathbf a_{\rm 2D}(x,y) \cdot \mathbf a_{\rm 3D}^{\rm proj}(x,y)$ = ${|\mathbf a_{\rm 2D}(x,y)||\mathbf a_{\rm
3D}^{\rm proj}(x,y)|}{\rm cos}\theta$.

Both angles and ratios are in good agreement. 
The angle between the two stays within 20$^{\circ}$, which is true for both snapshots.
By fitting Gaussians to the distributions, we find that the mean value of mass-weighted amplitude ratio distribution is $|\mathbf  a_{\rm 2D}(x,y) |/|\mathbf  a_{\rm 3D}(x,y)|$ =  2.36 $\pm$ 1.81 at the $t =  5.7 \times 10^5$ years snapshot.
For the simulation taken at the $t = 1.1 \times 10^6$ years, we find $|\mathbf  a_{\rm 2D}(x,y) |/|\mathbf  a_{\rm 3D}(x,y)|$ = 2.21 $\pm$ 1.78.
Gaussian fitting in the volume-weighed amplitude ratio distribution gives a mean value of 1.74 and 1.75 with a FWHM of 0.41 and 0.77 for two snapshots respectively.
These ratios can be understood as: when a 3D distance is projected onto 2D, the average distance ratio is <R$_{\rm 2D}$/R$_{\rm 3D}$> $\sim$ $2/\pi$. This may result in an average acceleration amplitude ratio of <${|\mathbf a_{\rm 2D}(x,y)|/|\mathbf a_{\rm 3D}^{\rm proj}(x,y)|}$> $\sim$ 2.47, which is similar to what is observed.

\subsection{Cause of significant deviations}
\begin{figure*}
\centering
\includegraphics[width=0.4\textwidth]{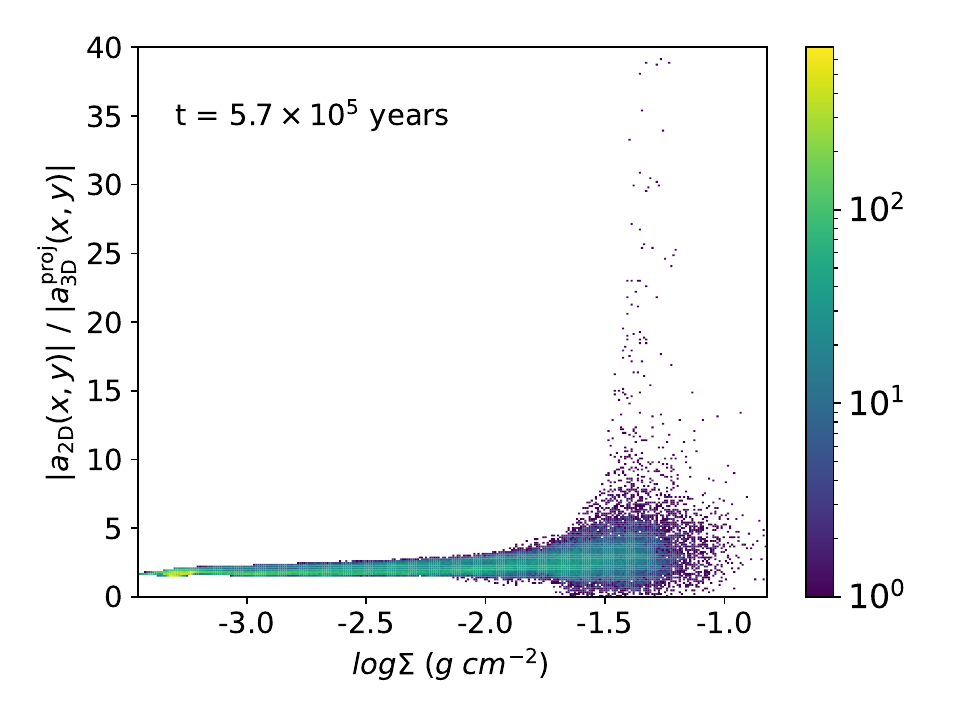}
\includegraphics[width=0.4\textwidth]{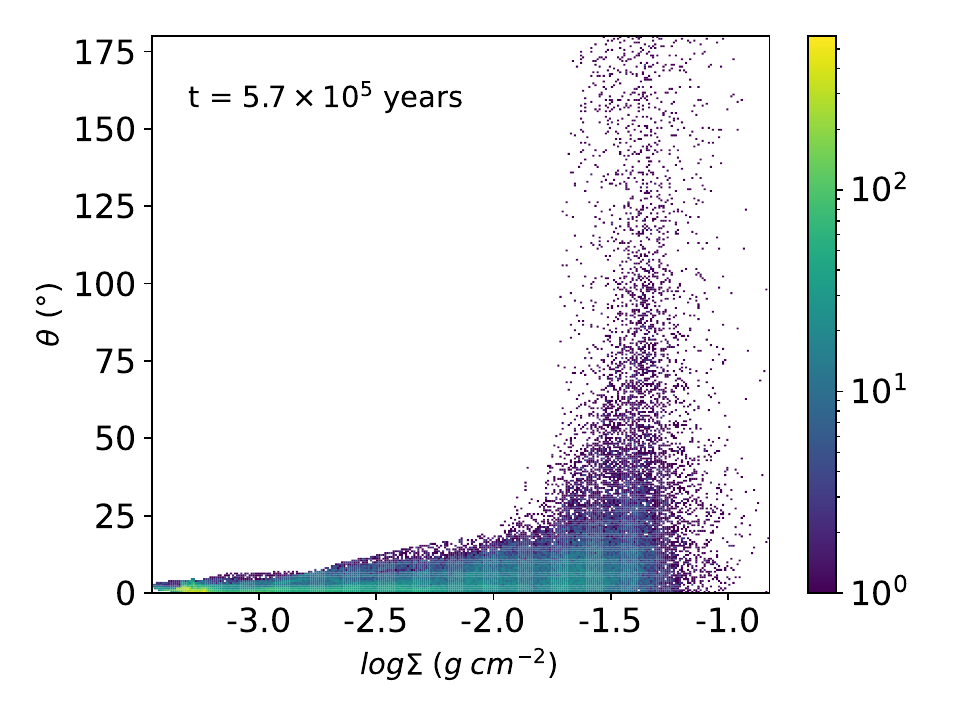}\\
\includegraphics[width=0.4\textwidth]{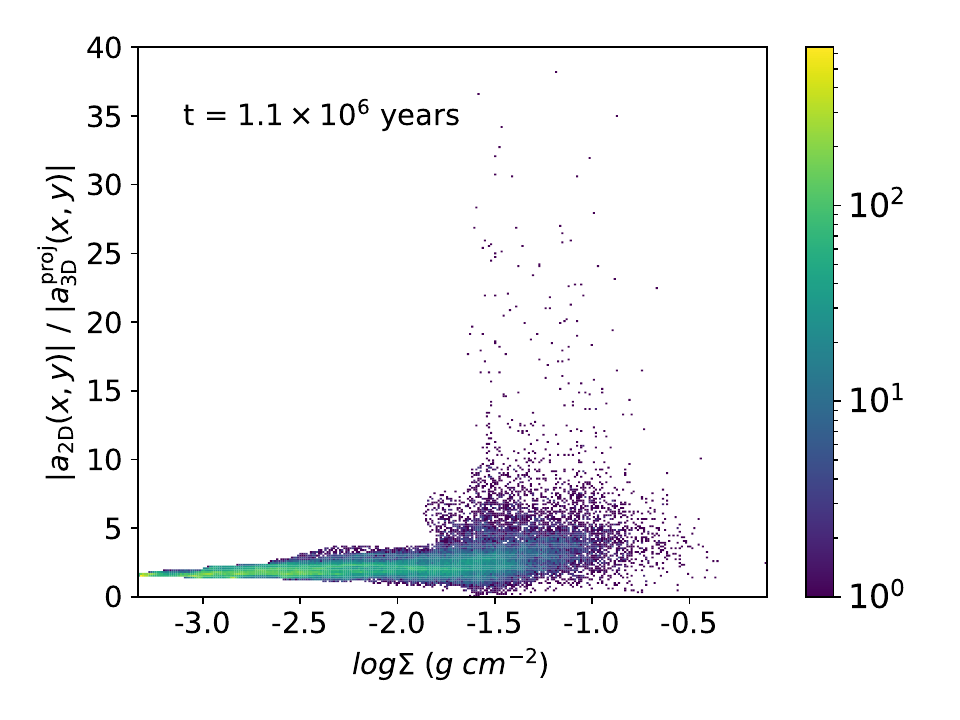}
\includegraphics[width=0.4\textwidth]{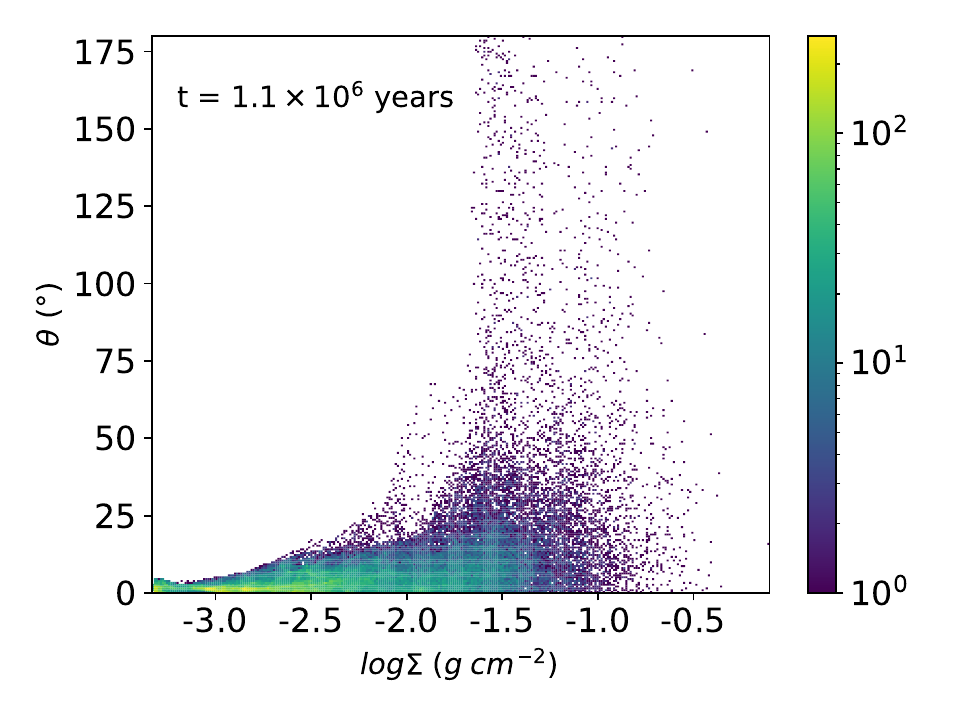}
\caption{The amplitude ratios ${|\mathbf a_{\rm 2D}(x,y)|/|\mathbf a_{\rm 3D}^{\rm proj}(x,y)|}$ and the include angles $\theta$ at different column densities. The error of both properties mainly occurs in places where the surface density $\Sigma \geq 10^{-1.5}$ g cm$^{-2}$.}\label{fig:ratiotheta}
\end{figure*}

Although the acceleration computed from the projected density distributions, in general, follows the real one, there are some significant deviations.
A first way to investigate these deviations is to plot amplitude ratios and included angles against the column density, as shown in Fig. \ref{fig:ratiotheta}.
Deviation mainly occurs in places where the surface density is relatively high ($\Sigma \ge 10^{-1.5}$ g cm$^{-2}$).

\begin{figure*}
\centering
\includegraphics[width=0.35\textwidth]{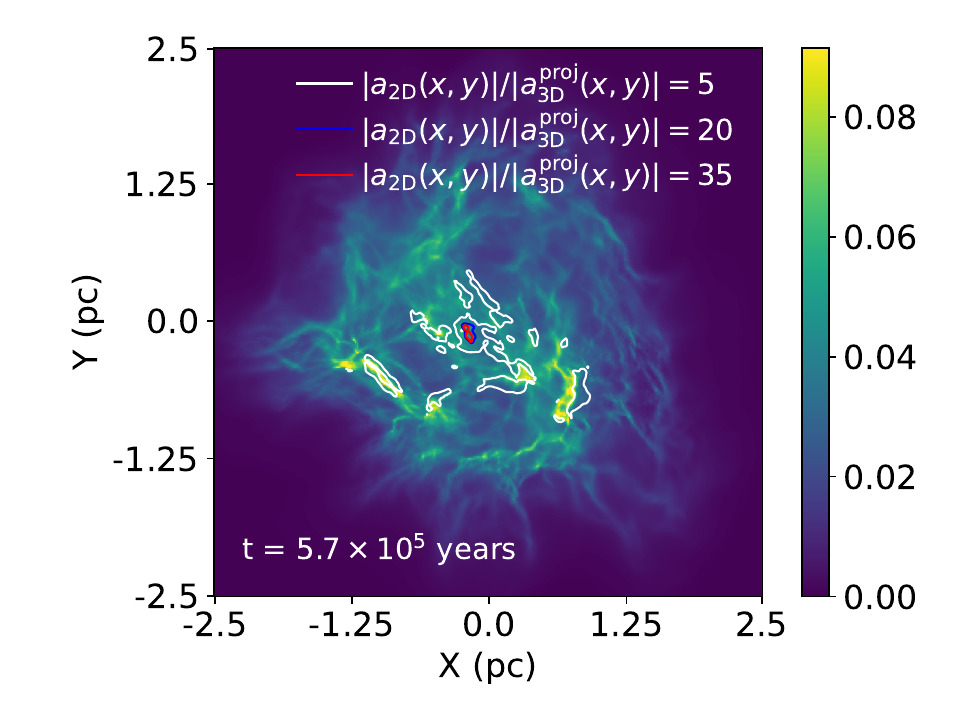}
\includegraphics[width=0.35\textwidth]{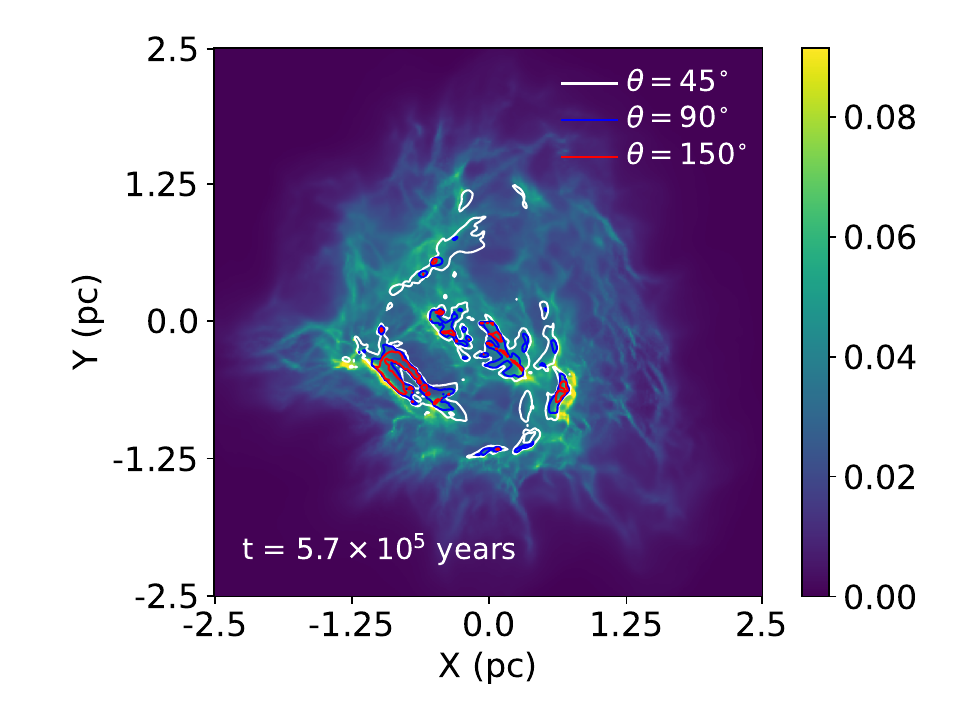}\\
\includegraphics[width=0.35\textwidth]{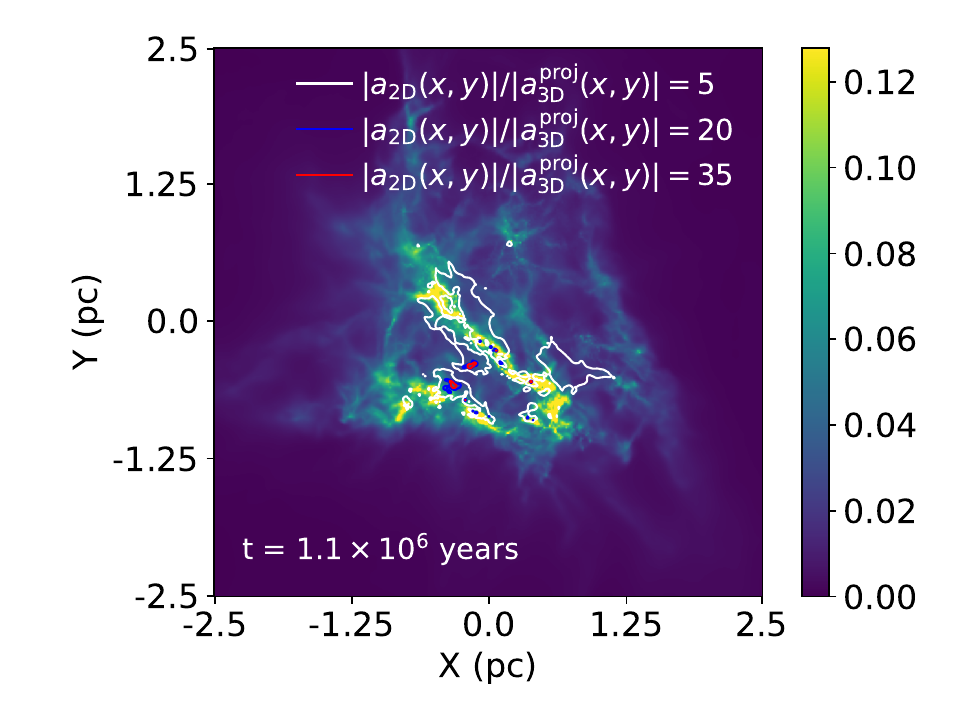}
\includegraphics[width=0.35\textwidth]{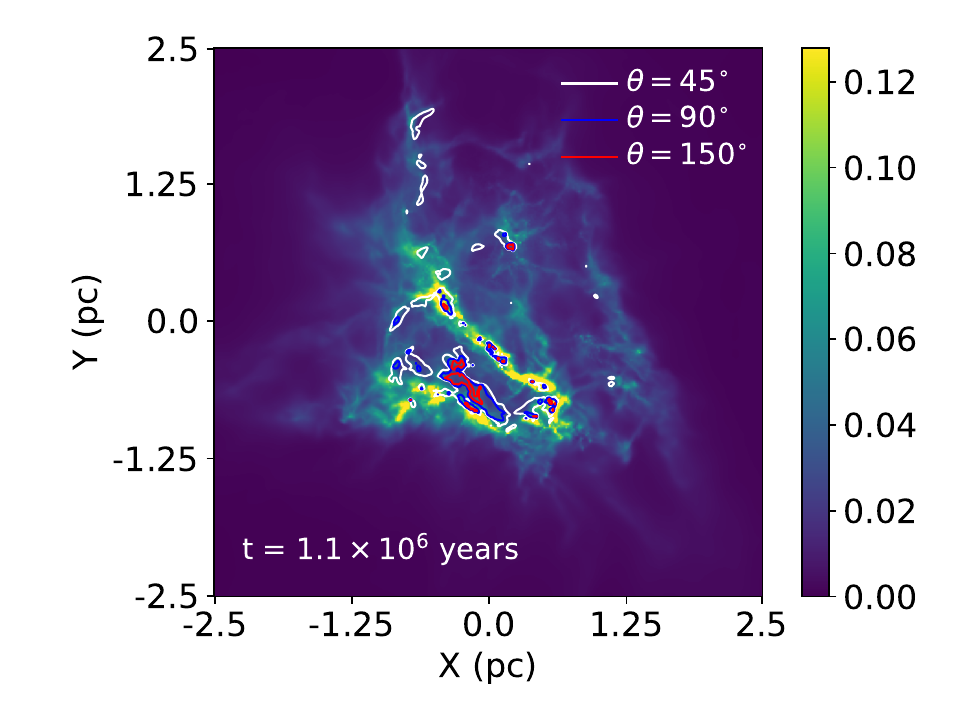}\\
\caption{Left two panel: The errors map of amplitude ratios ${|\mathbf a_{\rm 2D}(x,y)|/|\mathbf a_{\rm 3D}^{\rm proj}(x,y)|}$ overplotted onto surface density of clouds at different evolutionary stages. The white, blue and red contours represent positions with errors of 5, 20, and 35, respectively.
Right two panel: The errors map of include angles $\theta$ overplotted onto surface density of clouds at different evolutionary stages. The white, blue and red contours represent positions with errors of 45$^{\circ}$, 90$^{\circ}$, and 150$^{\circ}$, respectively.
}\label{fig:cont}
\end{figure*}

\begin{figure*}
\centering
\includegraphics[width=.35\textwidth]{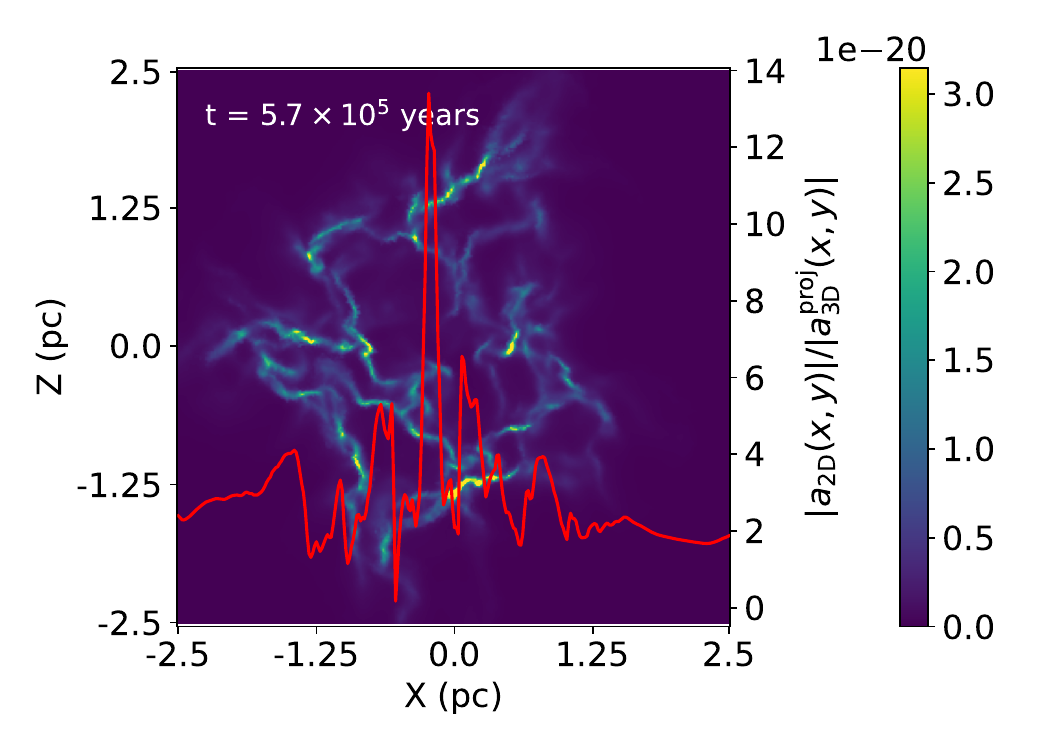}
\includegraphics[width=.35\textwidth]{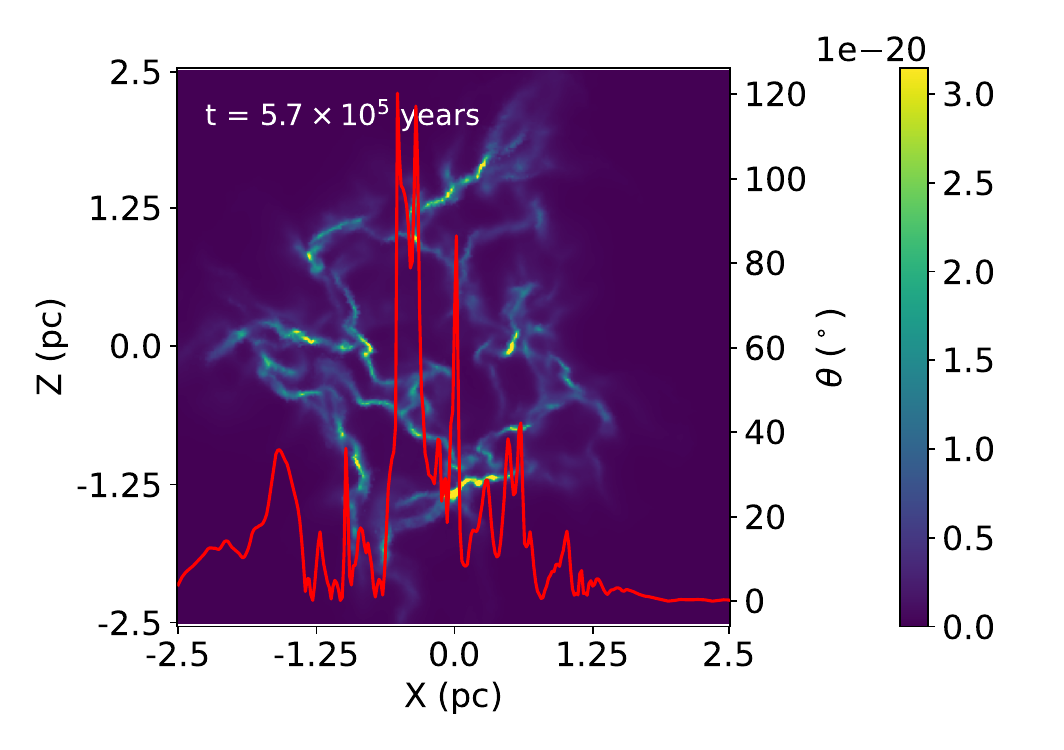}\\
\includegraphics[width=.35\textwidth]{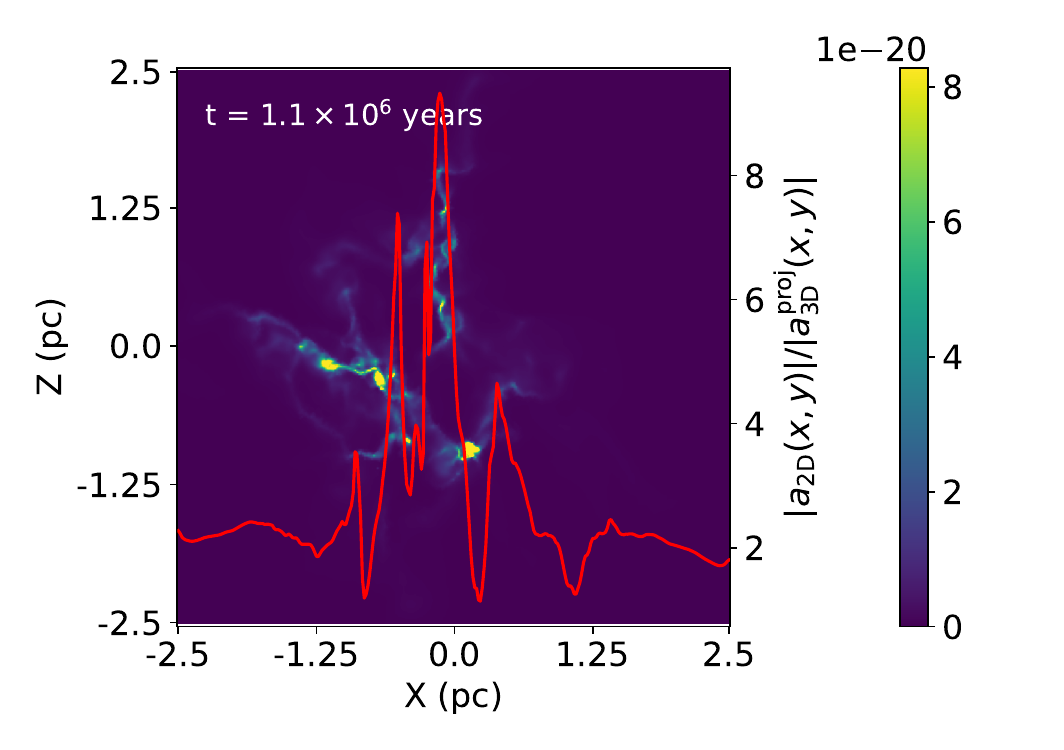}
\includegraphics[width=.35\textwidth]{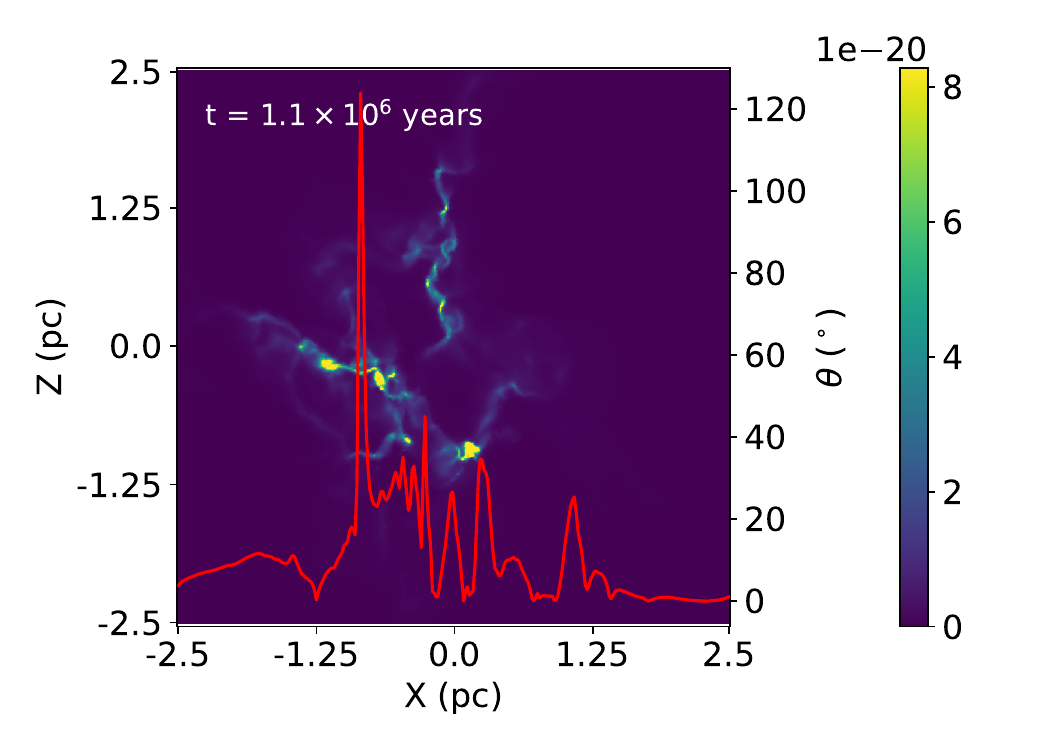}
\caption{Errors distribution of the amplitude ratios $|\mathbf a_{\rm 2D}(x,y)| / |\mathbf a_{\rm 3D}^{\rm proj}(x,y)|$ and the include angles $\theta$ along the line of sight. The background image is the density distribution of $x$ - $z$ plane at $y$ = 0.}\label{fig:x-z}
\end{figure*}

To investigate the origin of these errors, we identify regions where ${|\mathbf a_{\rm 2D}(x,y)|/|\mathbf a_{\rm 3D}^{\rm proj}(x,y)|} \ge 5$, as well as regions where $\theta \ge 45^{\circ}$ in Fig. \ref{fig:cont}.
We find that the large errors of both properties occur in diffuse regions surrounded by dense structures.
To illustrate this, we extract the density distribution along the $x-z$ plane at $y=0$,
where the amplitude and angle errors are plotted in Fig. \ref{fig:x-z}.
Large errors occur in the regions where complex structures are found along the line of sight.


By comparing results from simulations taken at $t = 5.7 \times 10^5$ years and $t = 1.1 \times 10^6$ years, we find that the acceleration map becomes slightly more accurate when the region is more evolved.
This is likely caused by the fact that gravity leads to centrally-condensed structures \citep{2022MNRAS.514L..16L}, which are simpler to resolve.

\section{Conclusion}\label{sec:con}
Using simulation data, we calculate the gravitational potential of real 3D density distributions and 2D density planes, then derive their acceleration maps ${\mathbf a_{\rm 3D}^{\rm proj}}$ (defined in Eq. \ref{eq}) and ${\mathbf a_{\rm 2D}}$, respectively.
By comparing these acceleration maps, we find that the acceleration maps computed in 2D are good approximations of the 3D gravitational acceleration field when viewed in projection.
The amplitude of the acceleration will be moderately over-estimated $\mathbf a_{\rm 2D}(x,y) \approx 2.3 \pm 1.8 \; \mathbf a_{\rm 3D}^{\rm proj}(x,y)$, and the angles between the two
stay within 20$^{\circ}$.
These significant errors result from overlapping structures viewed on the sky plane.
In general, the acceleration map computed from 2D provides a view of the 3D acceleration field that is reasonably accurate. We expect our technique to be useful in establishing the link between cloud morphology and gravitational collapse.

\section*{Acknowledgements}
GXL acknowledges supports from  
NSFC grant K204101220130, W820301904 and 12033005. This research was supported by the Excellence Cluster ORIGINS
which is funded by the Deutsche Forschungsgemeinschaft (DFG,
German Research Foundation) under Germany’s Excellence Strategy
- EXC-2094 - 390783311.\\

[He, ORCID: 0000-0001-7916-5614],
[Li, ORCID: 0000-0003-3144-1952],
[Burkert, ORCID: 0000-0001-6879-9822]

\bibliography{ref}
\bibliographystyle{mnras}
\end{document}